\documentclass{iopart}

\usepackage[utf8]{inputenc}
\usepackage{graphicx}
\usepackage[caption=false]{subfig}
\usepackage{iopams}
\usepackage{bm}
\usepackage{algorithm2e}
\usepackage{algorithmic}
\usepackage[        
 colorlinks=true,   
 citecolor=green,   
 linkcolor=cyan,    
 menucolor=blue,    
 urlcolor=magenta,  
 bookmarksopen=true,
 breaklinks
]{hyperref}

\newcommand{\nb}[1]{\partial #1}
\newcommand{\lld}{\Big\langle\!\!\Big\langle}
\newcommand{\rrd}{\Big\rangle\!\!\Big\rangle}

\newcommand{\avg}[1]{\langle #1 \rangle}
\newcommand{\prob}{P}

\newcommand{\sgn}{\textrm{sgn}}

\newcommand*{\argmax}{\textrm{arg}\,\textrm{max}}

\begin{document}

\paper{Statistical mechanics of reputation systems in autonomous networks}
\author{Andre Manoel$^1$ and Renato Vicente$^2$}
\address{
 $^1$ Dep. de Física Geral, 
 Instituto de Física,
 Universidade de São Paulo, 
 05314-090, São Paulo-SP, Brazil \\
 $^2$ Dept. of Applied Mathematics, 
 Instituto de Matemática e Estatística, 
 Universidade de São Paulo,
 05508-090, São Paulo-SP, Brazil}
\eads{\mailto{amanoel@if.usp.br}, \mailto{rvicente@ime.usp.br}}
\date{December 28th, 2012}

\begin{abstract}
  Reputation systems seek to infer which members of a community can be trusted
  based on ratings they issue about each other. We construct a Bayesian inference
  model and simulate approximate estimates using belief propagation (BP). The
  model is then mapped onto computing equilibrium properties of a spin glass in a
  random field  and  analyzed by employing the replica symmetric cavity approach.
  Having the fraction of trustful nodes and environment noise level as control
  parameters, we evaluate the theoretical performance in terms of estimation
  error and the robustness of the BP approximation in different scenarios.
  Regions of degraded performance are then explained by the convergence
  properties of the BP  algorithm and by  the emergence of a glassy phase.


\end{abstract}


\maketitle

\section{Introduction}
\label{sec:intro}

Ad-hoc \cite{Barbeau07} and wireless sensor networks \cite{Dargie10} work
in the absence of a central authority and are increasingly pervasive in
modern computer systems. The secure operation of these autonomous networks
depends on the capability of establishing trust among network entities. In
general it is reasonable to assume that reputation and trust are positively
correlated quantities and then employ a mutual scoring system as a source
of data that can be used to estimate reputations
\cite{Sabater2005,Mui2002a,Buchmann2007}.

Here we are concerned with the part of a reputation system \cite{Marmol2010}
that identifies ill-intentioned individuals or malfunctioning devices by
estimating reputations. This task would be trivial if the scores provided
a reliable representation for the reputation of an entity. Instead,
evaluation mistakes may happen or misleading ratings may be issued on
purpose \cite{Josang2009}.

Reputation systems are particularly prone to attacks by malicious entities
which can corrupt the recommendation process
\cite{Marmol2009,Hoffman2009,Josang2007}. This happens for instance when
multiple entities conspire to emit negative ratings about well-intentioned
agents while emitting positive ratings about co-conspirators. In another
form of attack, known as a \emph{Sybil attack}, a single entity could
impersonate others and trick the reputation mechanism.

The simplest algorithms employed by online communities use average ratings
to determine reputations. Despite having the advantage of being easy to
understand, these algorithms do not take into account the possibility of
entities committing mistakes or acting deceitfully, what often leads to
inferior results. More sophisticated algorithms employ Bayesian inference
\cite{Mui2002} or fuzzy logic \cite{Sabater2001}. Recently, iterative
formulas over looped or arbitrarily long chains -- the so called
\emph{flow models} -- have also been proposed (e.g. the PageRank algorithm
\cite{Page1999}). For a more thorough exposition of the range of techniques
employed we suggest recent reviews such as \cite{Sabater2005,Josang2007}.

In this paper, we employ statistical mechanics techniques to study the
performance of a belief propagation algorithm to approximately estimate
reputations. This analysis provides insights into the general structure of
the inference problem and suggests improvement directions.

The material is organized as follows. In section \ref{sec:model} we
introduce the inference model, an algorithm for estimating reputations and a
performance measure. In section \ref{sec:simulation} we simulate the
algorithm and discuss the results. A theoretical analysis is presented in
section \ref{sec:theor} and phase diagrams are calculated. Section \ref{sec:dynprop} discusses 
the dynamical properties of the approximate inference that impact performance. In section
\ref{sec:robust} we discuss the algorithm robustness by analyzing attacks,
parameter mismatches and different topologies. Finally, conclusions are
provided in section \ref{sec:concl}.

\section{Estimating Reputations}
\label{sec:model}

We model a reputation system along the lines of \cite{Ermon2009,Jiang2006}.
An entity $i=1,\cdots,n$ has a reputation $r_i$, and issues ratings $J_{ij}$
about other entities $j=1,\cdots,n$ with $j\neq i$. We define a set $\Omega$
of ordered pairs that contains $(i, j)$ if the rating that $i$ issues about
$j$ is present. We assume that ratings and reputations are related by a
given function
\begin{equation}
 \bm{J}= f (\bm{r}, \{\xi\}),
\end{equation}
where $\{\xi\}$ is a set of random variables representing externalities as,
for example, uncertainties affecting opinion formation or transmission. A
model is defined by specifying the domains of $r_i$ and $J_{ij}$, the
distribution of $\{\xi\}$ and the function $f$.

\begin{figure}[ht]
 \centering
 \includegraphics[width=0.25\textwidth]{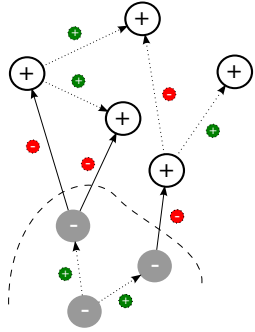}
 \caption{ {\bf Collusion:} agents of bad reputation issue positive ratings
 about each other and misguide the mechanism. Agents with good reputation
 are represented by white circles, and those of bad reputation by grey
 circles; arrow labels indicate whether a recommendation is positive
 (green) or negative (red).}
 \label{fig:collus}
\end{figure}

A good model should be able to describe realistic scenarios while
remaining amenable to analytical treatment. Ratings should represent true
reputations, namely, $J_{ij} \propto r_j$. The model should also take
into account emitter reputations, as we have to consider that an unreliable
entity may emit ratings defaming well-intentioned individuals or groups,
and that a sufficient number of such ratings can misguide the reputation
system -- the collusion phenomena depicted in figure~\ref{fig:collus}. A
simple choice is  
\begin{equation}
J_{ij} = \xi_{ij} r_i r_j, 
\label{eq:model}
\end{equation}
with $\xi_{ij}$ representing noise in the communication channel.  We start
by choosing $r_i, J_{ij},\xi_{ij} \in \{-1, 1\}$, with  $\xi_{ij}$ being a
random variable such that  $\xi_{ij} = 1$ with probability $p$ ({\it signal level}).

This choice of $f(\bm{r}, \{\xi\})$  mitigates the collusion
phenomena, as $J_{ij} = -1$ represents, in the noiseless case, either an
ill-intentioned positive recommendation about an unreliable entity or a
negative recommendation about a reliable entity issued by an unreliable
node. By introducing $\xi_{ij}$ we also build into the model misjudgments
and transmission failures. We assume a prior distribution for $\bm{r}$
supposing independence and a fraction $q$ of reliable agents ({\it
reputation bias}).

Our goal is to infer $\bm{r}$ given $\bm{J}$. For that we need the
posterior distribution $\prob(\bm{r} |\bm{J})$ that can be calculated with
help of Bayes theorem to find
\begin{eqnarray}
 \prob(\bm{r} | \bm{J}) &\propto& \prob(\bm{J} | \bm{r}) \prob(\bm{r})
 \label {eq:bayes} \\ &\propto& \prod_{(i, j) \in \Omega} \prob(\xi_{ij})
 \prod_i \prob(r_i). \nonumber
\end{eqnarray}

Notice that we can describe a  $\pm1$ random  variable with probability $p$
for $x=1$ by the distribution $\prob(x) \propto \exp (\alpha_p x)$, with
$\alpha_p = \frac{1}{2} \log \frac{p}{1-p}$. This yields
\begin{eqnarray}
 \prob(\bm{r} | J) &\propto& \prod_{(i, j) \in \Omega} \exp
 (\alpha_p \xi_{ij}) \prod_i \exp (\alpha_q r_i) \label{eq:post1} \\
 &\propto& \exp \left( \alpha_p \sum_{(i, j) \in \Omega} \xi_{ij} +
 \alpha_q \sum_i r_i \right) \\ 
 \label{eq:post}
 &\propto& \exp \left [\alpha_p \left( \sum_{(i, j) \in \Omega} J_{ij}
 r_i r_j + \frac{\alpha_q}{\alpha_p} \sum_i r_i \right) \right],
\end{eqnarray}
where $\alpha_q = \frac{1}{2} \log \frac{q}{1-q}$.

Alternatively, the dependence structure of the inference model may be
represented by a directed graph $\mathcal{G} = (V, E)$, with each vertex $v
\in V$ standing for an entity, and an arc $(i, j)$ being connected if and
only if $(i, j) \in \Omega$. 

The methods of equilibrium statistical mechanics require a symmetric
$\bm{J}$ which can be achieved by grouping terms as \[\sum_{(i, j) \in
\Omega} J_{ij} r_i r_j = \frac{1}{2} \sum_{ i j} (J_{ij} + J_{ji}) r_i
r_j,\] where in the r.h.s. we assume that $J_{ij}=0$ if $(i, j) \notin
\Omega$. We here only consider undirected graphs, to say, both arcs $(i, j)$
and $(j, i)$ are in $\mathcal{G}$ if $(i,j)\in \Omega$.  By replacing
$\frac{1}{2} (J_{ij} + J_{ji})$ for $J_{ij}$, we work with $J_{ij} \in \{-1,
0, 1\}$, discarding dissonant opinions that, fortunately, are few in many
cases of interest \cite{Josang2007}. The case where a single arc may be
present can be modeled by a simple extension that considers $J_{ij} \in
\{-1, -\frac{1}{2}, 0, \frac{1}{2},1\}$.

Given a sample of ratings $\{J_{ij}\}$, our goal is to find an estimate
$\bm{\hat{r}}$ for the reputations $\bm{r}$ while keeping the error 
\begin{equation}
 \varepsilon (\bm{\hat{r}}, \bm{r}) = \frac{1}{2} \left(1 -
 \frac{\bm{\hat{r}} \cdot \bm{r}}{n} \right), \label{eq:erro}
\end{equation}
as small as possible.

A naive solution calculates reputations according to the majority of
recommendations (a {\it majority rule}) 
\begin{equation}
 \hat{r}_i = \sgn \left( \sum_{k \in \nb{i}} J_{ki} \right),
 \label{eq:triv}
\end{equation}
where $\partial i$ stands for the neighborhood of the vertex $i$ in
$\mathcal{G}$. Assuming that ratings are produced according to eq.
\ref{eq:model} we can write
\begin{equation}
  \hat{r}_i = \sgn \left[ \left( \sum_{k \in \nb{i}} \xi_{ki} r_k \right)
  r_i \right].
\end{equation}

Thus we can calculate the probability of agreement between $\hat{r}_i$ and
the real reputation $r_i$ by considering that $r_k$ and $\xi_{ki}$ are
random variables sampled from $\prob(r)$ and $\prob(\xi)$ (while keeping
$\xi_{ik} = \xi_{ki}$), and by introducing a random variable $\lambda$
sampled from the degree distribution of graph $\mathcal{G}$: 
\begin{eqnarray}
 \prob(\hat{r}_i = r_i) &=& \lld P\left( \sum_{k = 1}^\lambda \xi_{ik} r_k > 0
 \right)\rrd_{\lambda}\nonumber\\ 
 &=&  \lld \sum_{n = 0}^{\lfloor \frac{\lambda - 1}{2} \rfloor}
 {\lambda \choose n} \varpi^{(\lambda-n)} (1 - \varpi)^n \rrd_{\lambda},
 \label{eq:triv_pred}
\end{eqnarray}
with $\varpi = pq + (1-p)(1-q)$, and $\lld \cdot \rrd_\lambda$ denoting the
average with respect to $\lambda$. Here, $\xi_{ik} r_k$ is a $\pm 1$ random variable
with parameter $\varpi$ (i.e., it is $+1$ with probability
$\varpi$).  The sum of a set of these variables is binomially distributed,
and the expression in the r.h.s. represents the cumulative distribution.

Until section \ref{sec:robust} we assume a random regular graph
$\mathcal{G}$ with degree $c$. Also, we deal with a scenario such that
signal prevails $p > 0.5$ and most of the nodes are reliable $q > 0.5$.

\begin{figure}[ht]
 \centering
 \includegraphics[width=0.3\textwidth]{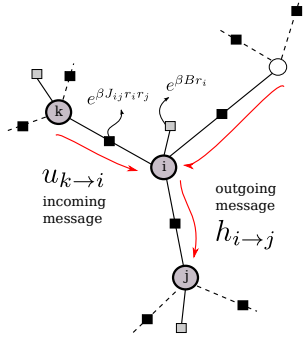}
 \caption{{\bf Factor graph and BP message-passing.} Snapshot of a factor
 graph representing  the posterior \ref{eq:posteriorBP}. Variable nodes are
 represented by circles, factor nodes are represented by small squares.  At
 an iteration outgoing messages $h_{i \to j}^{(t+1)}$ are computed after
 combining incoming messages $u_{k \to i}^{(t)} $. When convergence is
 attained we compute effective fields $\hat{h}_i$ and the approximation for
 marginal posteriors $\prob(r_i)$.}
 \label{fig:bp}
\end{figure}

For an entity $i$ the inference task consists of maximizing the posterior
for $\bm{r}$ marginalized over every component except $i$. The posterior is
factorizable and can be put in the following form
\begin{equation}
 \prob(\bm{r} | \bm{J}) = \frac{1}{\mathcal{Z}} \prod_{(i, j)\in
   \mathcal{G}} \exp (\beta J_{ij} r_i r_j) \prod_i \exp (\beta B r_i),
\label{eq:posteriorBP}
\end{equation}
where $\beta$ and $B$ are parameters which are optimally set at Nishimori's
condition \cite{Iba1999} $\beta=\alpha_q$ and $B=\frac{\alpha_q}{\alpha_p}$.
Factorizable distributions such as this are well represented by factor
graphs \cite{Kschischang01,Loeliger2004,Mezard2009}, with variable nodes
associated to the $\{r_i\}$ and function (or factor) nodes representing the
functions linking them, in this case the exponentials. Figure \ref{fig:bp}
zooms in a factor graph representation of the posterior
\ref{eq:posteriorBP}.

Given this posterior distribution we calculate marginal distributions
$\prob(r_i) = \sum_{r_j \neq r_i} \prob(\bm{r} | \bm{J})$ efficiently by
employing the message-passing scheme of belief propagation (BP) on the
factor graph associated to the posterior eq. \ref{eq:posteriorBP}
\cite{Mezard2009}. In our case the outgoing BP messages (see figure
\ref{fig:bp} for illustration) are 
\begin{equation}
 h_{i \to j}^{(t+1)} = B + \sum_{k \in \nb{i} / j} u_{k \to i}^{(t)}
 (J_{ki}, h^{(t)}_{k \to i}),
 \label{eq:BPeq}
\end{equation}
while $u_{k \to i}^{(t)} = \frac{1}{\beta} \tanh^{-1} \left[ \tanh (\beta
J_{ki}) \tanh (\beta h_{k \to i}^{(t)}) \right]$ are incoming messages. 

Iterating this set of equations until convergence we obtain $\{h_{i \to
j}^\ast, u_{i \to j}^\ast\}$ that yields an approximation for marginals
$\prob(r_i) \propto \exp (\beta \hat{h}_i r_i)$, with effective fields
given by
\begin{equation}
\hat{h}_i = B + \sum_{k \in \nb{i}} u_{k \to i}^\ast.
\end{equation}
This algorithm is exact on trees, but can be used in graphs of any topology,
leading to good approximations provided that the average cycle length is
large \cite{Mezard2009}.

\begin{figure}[ht]
 \centering
 \includegraphics[width=0.5\textwidth]{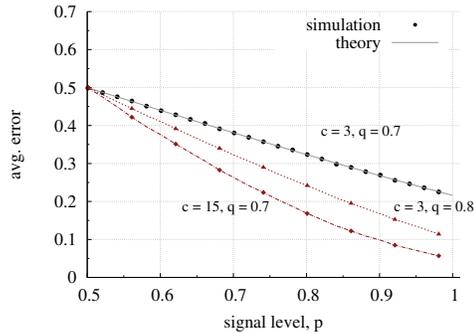}
 \caption{{\bf Estimation error for the majority rule.} Average errors in
 simulations (circles) are compared with theoretical errors calculated by
 averaging eq.~\ref{eq:erro} over \ref{eq:triv_pred} (lines). Averages are
 over 3000 runs, $q = \prob(r_i = 1) = 0.7$ or $q=0.8$. Graphs
 $\cal G$ are random and regular with degree $c = 3$ and $n = 100$.  Error
 bars are smaller than symbols in all three curves.}
\label{fig:triv}
\end{figure}

After convergence of the BP scheme reputations $r_i$ are estimated by
marginal posterior maximization (MPM) 
\begin{equation}
 \hat{r}_i = \argmax_{r_i} \prob(r_i) = \sgn(\hat{h}_i).
 \label{eq:prop}
\end{equation}

\section{Simulations}
\label{sec:simulation}

As a basis of comparison we run the majority rule algorithm defined by
eq.~\ref{eq:triv} for 3000 scenarios with reputations $\bm{r}$ chosen
randomly with  reputation bias $q$ and symmetric $\xi_{ik}$ with signal
level $p$. In figure \ref{fig:triv} we compare the average error in
simulations of the majority rule with the theoretical error calculated by
averaging eq.~\ref{eq:erro} over the distribution \ref{eq:triv_pred}. The
majority rule is not very far from what is used in common reputation systems
on e-commerce websites.  Note, however, that the error of this very simple
scheme can be larger than $1-q$ if the signal level $p$ is low enough.  The
message is therefore clear: in noisy environments assigning good reputations
by default  may be actually  more effective than using the majority
recommendation.

\begin{figure}[ht]
 \centering
 \includegraphics[width=0.5\textwidth]{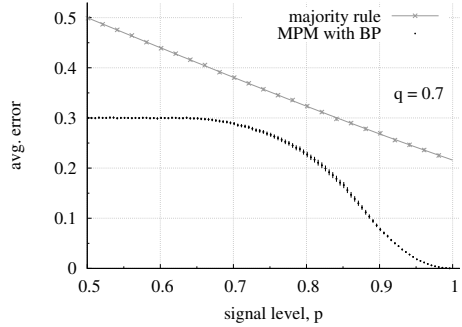}
 \caption{{\bf MPM estimate versus the majority rule. } Comparison between
 the average error of the majority rule (eq. \ref{eq:triv}) and of the MPM
 estimate (eq. \ref{eq:prop}), for $q \equiv \prob(r_i = 1) = 0.7$, a
 regular random graph with degree $c = 3$ and $n = 100$. For the majority
 rule we show theory and simulations, with symbols larger than error bars.
 For the MPM estimate we show the average over $3000$ simulation runs with
 error bars representing one standard deviation. } 
 \label{fig:triv_compar}
\end{figure} 

In figure \ref{fig:triv_compar} we compare the majority rule with the average
over $3000$ runs of the MPM estimate computed with the BP algorithm. For
convenience, the algorithm is presented as a pseudocode in the \ref{sec:mpm}
\footnote{Source code is also available at
  \url{https://github.com/amanoel/repsys}.}. The gains in performance when
the collusion phenomena is built into the inference model are considerable
even in very noisy environments. 

\begin{figure}[ht]
 \centering
 \subfloat[average error]{
 \includegraphics[width=0.37\textwidth]{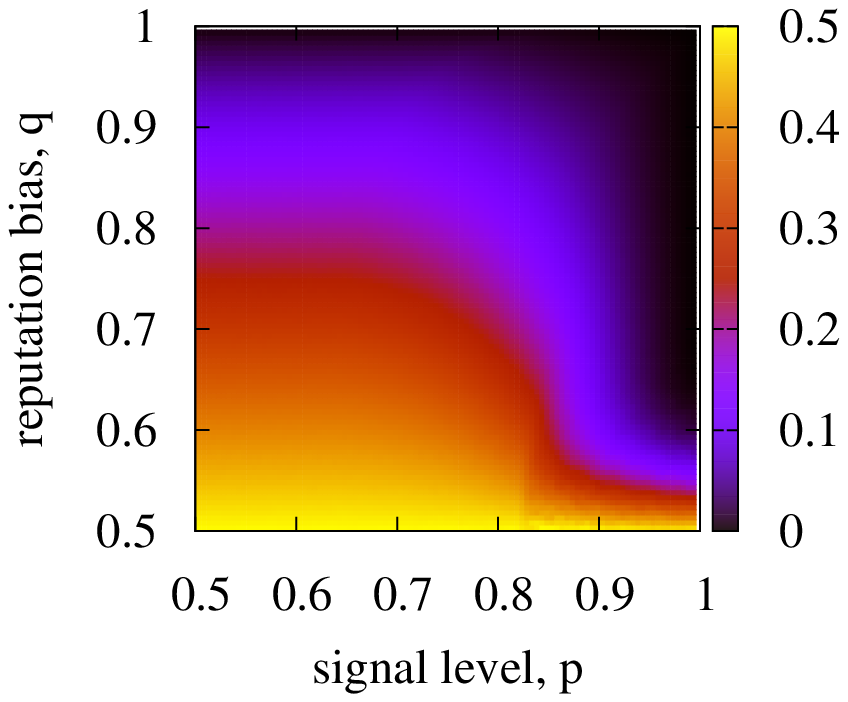}
 \label{fig:simul_error}
 }
 \subfloat[iterations to convergence]{
 \includegraphics[width=0.37\textwidth]{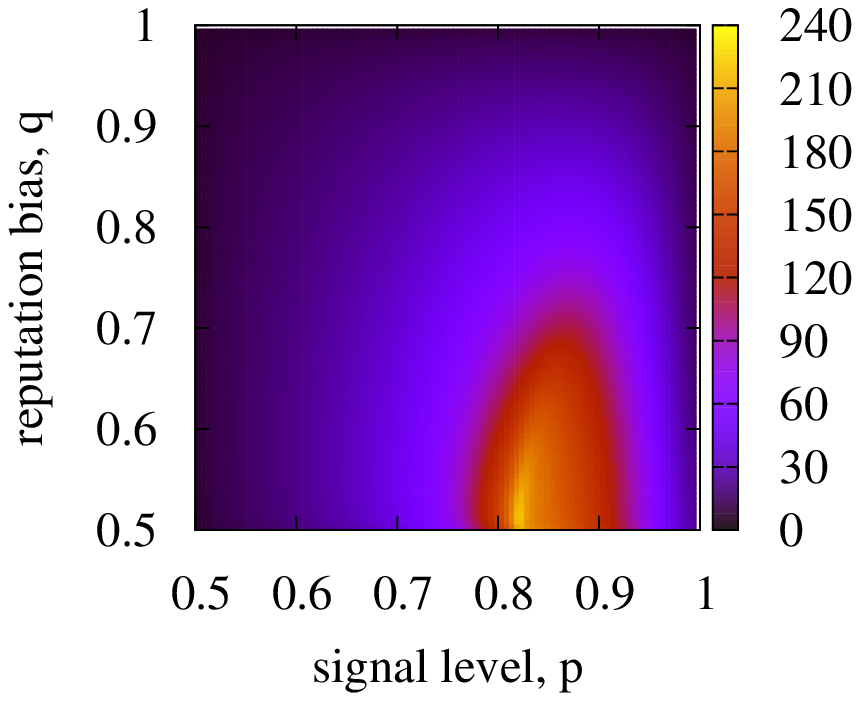}
 \label{fig:simul_convrate}
 }
 \caption{ {\bf Average error and number of iterations to convergence:}
   $3000$ runs of the BP algorithm in a grid of values for signal level $p$
   and reputation bias $q$, with $n = 100$ and $c = 3$. In each run
   $\mathcal{G}$, $\{r_i\}$ and $\{\xi_{ij}\}$ are sampled and ratings
   $\{J_{ij}=\xi_{ij}r_i r_j\}$ are calculated. Panel (a) depicts the
   average error as in eq.~\ref{eq:erro}. Panel (b) represents the average
   number of iterations it took for the BP algorithm to converge.}
\end{figure}

A detailed view of the error surface for the approximate MPM estimates in
terms of reputation bias $q$ and signal level $p$ is depicted in figure
\ref{fig:simul_error}.  Two regions can be discerned with large error for
low signal level (high noise) and low reputation bias.  Other average
quantities can also be evaluated in the simulations in order to access the
algorithm's performance. Figure \ref{fig:simul_convrate}, for instance,
shows the average number of iterations it took until convergence has been
achieved. A distinctive region is observed with degraded time to
convergence.

The inference algorithm has to assume (or estimate) values $\hat{p}$ and
$\hat{q}$ for the signal level and reputation bias. Ideally 
parameters have to be set to the same values used to generate data.
However, the environment can change without warning and we would also like
to know how the inference scheme would perform in such circumstances. By
simulation we can generate the vector $\bm{r}$ and the symmetric matrix
$\xi$ as random variables with probability of being $+1$ set to $q$ and $p$,
respectively, and then run the algorithm assuming $\alpha_{\hat{p}}$ and
$\alpha_{\hat{q}}$.

\begin{figure}[ht]
 \centering
 \subfloat[empirical error]{
 \includegraphics[width=0.37\textwidth]{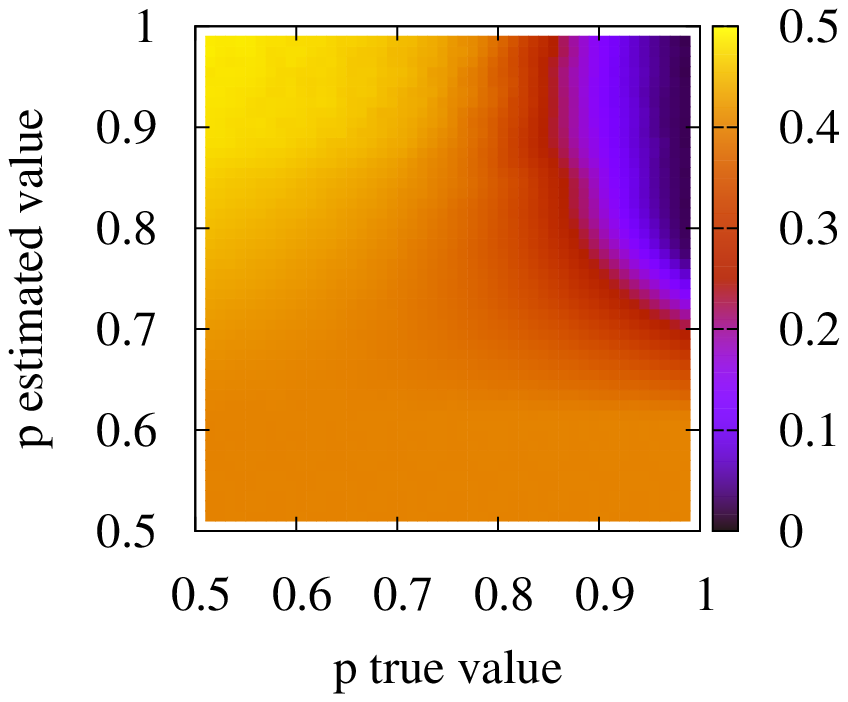}
 \label{fig:ppHat_error}
 }
 \subfloat[iterations to convergence]{
 \includegraphics[width=0.37\textwidth]{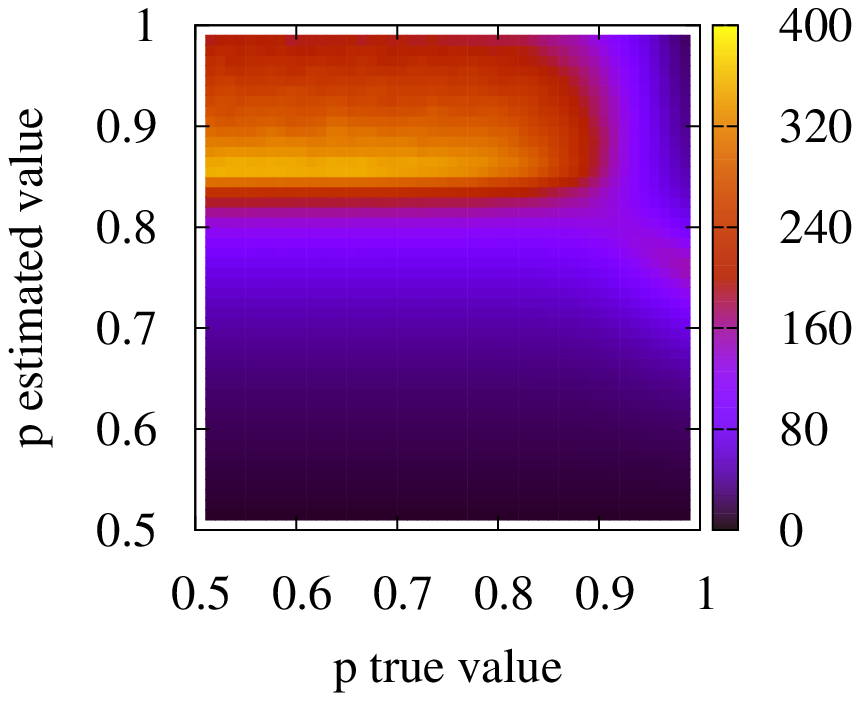}
 \label{fig:ppHat_convrate}
 }
 \caption{{\bf Empirical analysis for mismatched signal level $p$ with
   reputation bias $q = 0.6$}. Panel (a): empirical error; panel (b): average number of 
   iterations to convergence. Note that the performance is sensitive to small
   parameter mismatches in the neighborhood of $p=0.85$.}
 \label{fig:ppHat}
\end{figure}

Figure \ref{fig:ppHat} depicts results of this simulations in the plane
$p$--$\hat{p}$ for $q = 0.6$. Small mismatches between $\hat{p}$ and $p$ are
in general well absorbed by the inference scheme. A quick inspection of
figure \ref{fig:ppHat_diag} reveals that for $q = 0.6$, the neighborhood of $p
= 0.85$ exemplifies a specially sensitive region. Performance also
deteriorates when estimates, in this case for the signal level, are too
optimistic. 

In the next section we use equilibrium statistical mechanics to calculate
the phase diagram as a function of the control parameters $p-q$ and
$p-\hat{p}$, and to explain low performance regions in both cases.

\section{Theoretical analysis}
\label{sec:theor}

The posterior distribution \ref{eq:post} suggests the description of the
problem of inferring reputations in terms of the equilibrium properties of a
spin glass in an external field
\begin{equation}
 \mathcal{H} (\bm{s}|\bm{J},B) = -\sum_{(i, j) \in \Omega} J_{ij} s_i s_j -
 B \sum_i s_i,
 \label{eq:hamilt}
\end{equation}
where we designate the dynamic variable as $s_i$, and the \emph{target}
variables $r_i$ are fixed (\emph{quenched}). The MPM estimates described in
section \ref{sec:simulation} correspond to $\hat{r}_i = \sgn (m_i) $, with
$m_i = \tanh (\beta \hat{h}_i)$ representing local equilibrium
magnetizations. 

At Nishimori's condition \cite{Iba1999} temperature and field are chosen as
$\beta_N = \alpha_p$ and $B_N = \frac{\alpha_q}{\alpha_p}$, and the
microstates of this physical system are distributed according to a Gibbs
measure 
\begin{equation}
  \prob(\bm{s} | \bm{J})= \frac{1}{\mathcal{Z}} e^{-\beta_N
    \mathcal{H}(\bm{s}|\bm{J},B_N) }
\end{equation}
Other values of $\beta$ and $B$, corresponding to misspecified $p$ and $q$
can also be studied along the same lines.

We wish to calculate the equilibrium average error of eq.~\ref{eq:erro}
which corresponds to the magnetization of the Hamiltonian in eq.
\ref{eq:hamilt} (gauge) transformed with $r_i s_i \to s_i$. As this
Hamiltonian is not gauge invariant we now have to deal with a spin glass in
a random field

\begin{equation}
 \mathcal{H} (\bm{s}|\bm{\xi},\bm{r},B) = -\sum_{(i, j) \in \Omega} \xi_{ij}
 s_i s_j - B \sum_i r_i s_i.
 \label{eq:hamilt_rot}
\end{equation}

Here $\{\xi_{ij}\}$ and $\{r_i\}$ are quenched variables with
$\prob(\xi_{ij}=\xi_{ji}=1) = p$ and $\prob(r_i = 1) = q$. The error
$\varepsilon (\bm{\hat{r}}, \bm{r})$ in the gauge transformed variables can
be written as $\bar{\varepsilon} = \frac{1}{2} (1 - \langle \sgn(m)
\rangle)$, where $m = \frac{1}{n} \sum_i s_i$ is the gauge transformed
equilibrium magnetization. 

In our analysis we employ the replica-symmetric cavity method  along the
lines of \cite{Mezard2009,Matsuda2011}. In this section we calculate the
phase diagram at Nishimori's condition. We first write an equation for the
distribution of cavity fields for the gauge transformed variables as
calculated by the BP procedure in eq. \ref{eq:BPeq}

\begin{equation} 
 \prob(h) = \int \prod_{i=1}^{c-1} dh_i \prob(h_i)\,\lld \delta \left(h - Br
 - \sum_{i = 1}^{c-1} u_i(\xi_i,h_i)\right)\rrd_{r, \{\xi_i\}},
 \label{eq:cav-rot_int}
\end{equation}
or, more concisely

\begin{equation}
 h \stackrel{d}{=} \lld Br + \sum_{i = 1}^{c-1} u_i(\xi_i, h_i) \rrd_{r,
   \{\xi_i\}},
 \label{eq:cav-rot}
\end{equation}
with $\stackrel{d}{=}$ indicating equality in distribution. In this context,
$h$ is the cavity field, and 

\begin{equation*}
 u_i(\xi_i,h_i) = \frac{1}{\beta} \tanh^{-1} \left[ \tanh (\beta \xi_i)
 \tanh (\beta h_i) \right]
\end{equation*}
are cavity biases \cite{Mezard2009,Zdeborova2009}. Note that we work here
under the assumption that it makes sense to describe fixed points of the BP
equations \ref{eq:BPeq} in terms of a unique density $\prob(h)$. That is 
the replica symmetry (RS) assumption. 
 
We calculate numerical solutions to \ref{eq:cav-rot_int} by the population
dynamics algorithm (see \ref{sec:popdyn} for details) and then
calculate thermodynamic quantities. The magnetization is given by $m =
\langle \tanh (\beta \hat{h}) \rangle_{\hat{h}}$, where 
\begin{equation} 
  \hat{h} \stackrel{d}{=} \lld Br + \sum_{i = 1}^{c} u_i (\xi_i, h_i)
  \rrd_{r, \{\xi_i\}}
\end{equation}
is the effective field. Note that the sum here ranges from $1$ to $c$.

The onset of a spin glass phase can be detected by finding divergences in
the spin glass susceptibility. Provided that the disorder is spatially
homogeneous, the spin glass susceptibility averaged over this disorder can
be written as:
\begin{equation}
  \chi_{sg}=\sum_{\ell=0}^{\infty} \mathcal{N} (\ell) \lld \left[ \avg{s_0
  s_\ell} - \avg{s_0} \avg{s_\ell} \right]^2\rrd_{r,\{\xi_i\}, \mathcal{G}}.
\end{equation}
where $s_0$ is a variable at an arbitrary central site, $s_\ell$ is an
arbitrary variable at a site separated from $0$ by a chemical distance
$\ell$ and $\mathcal{N} (\ell)$ is the number of sites at a distance $\ell$
from $0$. 

The fluctuation-dissipation theorem, the symmetry introduced by averaging
and the BP equations \ref{eq:BPeq} yield

\begin{equation}
  \chi_{sg} = \sum_{\ell=0}^{\infty} \mathcal{N} (\ell) \lld \left[
  \frac{\partial m_0}{\partial \hat{h}_{0}} \frac{\partial
  \hat{h}_{0}}{\partial u_{\cdot \to 0}} \frac{\partial u_{\cdot \to
  0}}{\partial u_{\cdot \to \ell}} \frac{\partial u_{\cdot \to
  \ell}}{\partial \hat{h}_{\ell}} \right]^2\rrd_{r, \{\xi_i\},
  \mathcal{G}},
\end{equation}

where $u_{\cdot \to 0}$ and $u_{\cdot \to \ell}$ represent incoming messages
in a path connecting $0$ to $\ell$. A sufficient condition for $\chi_{sg}$
to diverge is, therefore, that 

\begin{equation}
  \lim_{\ell \to
  \infty} \left[\frac{\partial u_{\cdot \to 0}}{\partial u_{\cdot \to
  \ell}}\right]^2 >0.
\end{equation}

This quantity measures the sensibility of the incoming message at a central
site $0$ to a perturbation in a message forming at a  far outside  distance
$\ell$. In terms of cavity field distributions we can write

\begin{equation}
  \rho = \lld\left\langle \lim_{\ell\to \infty} \left[\frac{\partial
    u_{0}[u_\ell(h)]}{\partial u_{\ell}}\right]^2 \right\rangle_h\rrd_{r,
      \{\xi_i\}}.
\end{equation}
 
A number of numerical methods can be used to evaluate $\rho$
\cite{Zdeborova2009,Matsuda2011}. Using population dynamics (described in
\ref{sec:popdyn}), we introduce two slightly different initial states such
that $\bm{u'}_0[i] - \bm{u}_0[i] = \delta$ with $i=1,\cdots,N$ and $\delta =
10^{-4}$. After a large number $\tau$ of iterations of the population
dynamics algorithm we calculate 

\begin{equation}
  \rho \approx \frac{1}{N} \sum_i (\bm{u'}_\tau[i] - \bm{u}_\tau[i])^2. 
\end{equation}

The order parameters $m$ and $\rho$ allow the identification of four
different thermodynamic phases: paramagnetic if $\rho = 0$, $m = 0$,
ferromagnetic if $\rho = 0$, $m > 0$, glassy if $\rho \neq 0$, $m = 0$
and, mixed or ferromagnetic ordered spin glass for $\rho \neq 0$, $m >
0$. 

\begin{figure}[ht]
 \centering
 \includegraphics[width=0.5\textwidth]{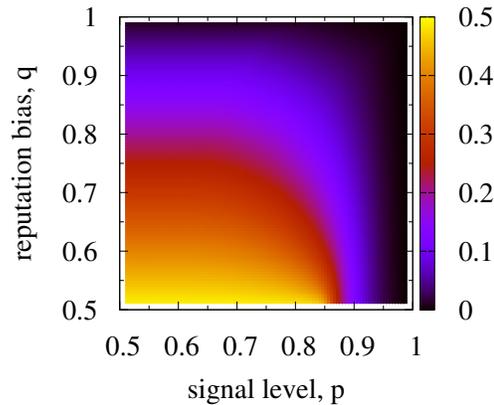}
 \caption{{\bf Theoretical error at Nishimori's
 condition:} theoretical error $\bar{\varepsilon} = \frac{1}{2} \left( 1
 - \langle \sgn (\hat{h}) \rangle_{\hat{h}} \right)$ for different values of
 $p$ and $q$.}
 \label{fig:nish_error}
\end{figure}

To measure the theoretical error, we also calculate $\langle \sgn (\hat{h})
\rangle_{\hat{h}}$.  The theoretical results depicted in figure
\ref{fig:nish_error} are corroborated by simulations depicted in figure
\ref{fig:simul_error} .  Under Nishimori's condition, $\rho = 0$ for all
values of $p$ and $q$, and thus no glassy or mixed phase is present.
Likewise, for all values of $p, q > 0.5$, $m > 0$, and the ferromagnetic
phase covers the whole region.

Three rigorous results on phase diagrams of similar models are available:
(A) if the Hamiltonian is gauge invariant the Nishimori line does not
cross a spin glass phase  \cite{NishimoriBook}; (B) there is no spin
glass phase in a random field Ising model \cite{Krzakala10}; and (C)
provided that the parameters employed in the inference task are identical to
those used to generate data (namely, we are at Nishimori's condition), in a
random graph with bounded maximum degree the BP scheme converges to the
correct marginals in the thermodynamic limit \cite{Montanari2008}.  

\begin{figure}
 \centering
 \subfloat[theoretical error]{
 \includegraphics[width=0.37\textwidth]{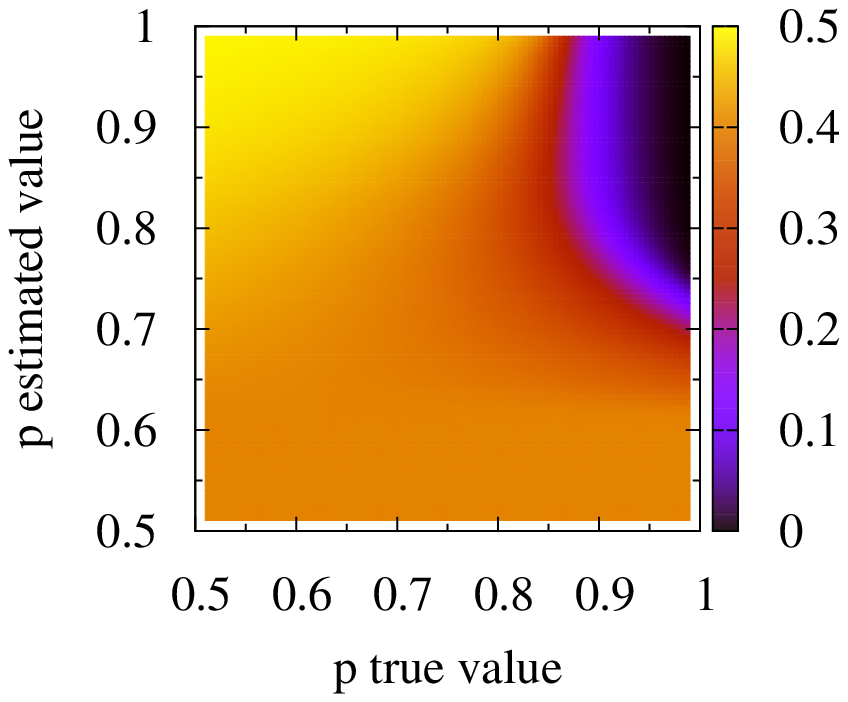}
 \label{fig:ppHat_therror}
 }
 \subfloat[phase diagram]{
 \includegraphics[width=0.37\textwidth]{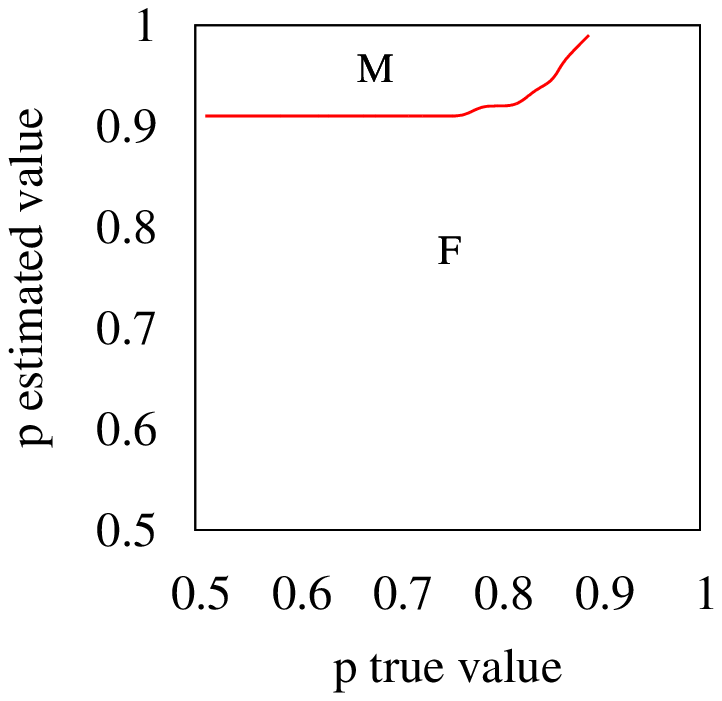}
 \label{fig:ppHat_diag}
 }\\
 \subfloat[$\rho$ as a function of $\hat{p}$]{
 \includegraphics[width=0.37\textwidth]{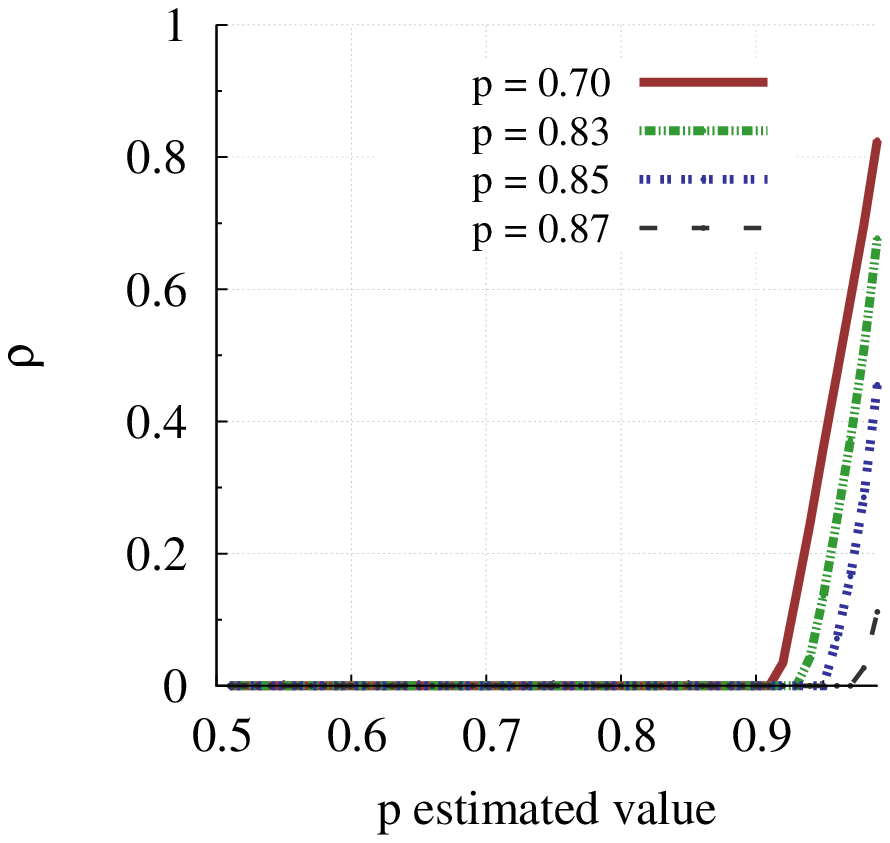}
 \label{fig:ppHat_rho}
 }
 \caption{{\bf Theoretical analysis for mismatched signal level $p$ with
   reputation bias $q = 0.6$}. Panel (a): theoretical error, in good agreement
   with simulations. Panel (b): phase diagram, in which a
   mixed phase appears, explaining  the algorithm's  degraded  performance 
   in that region. Panel (c): $\rho$ signals the onset of the mixed phase as 
   $\hat{p}$ increases.}
 \label{fig:ppHat_theor}
\end{figure}

In order to check our results we observe that choosing $q=1/2$
($\alpha_q=0$) yields a gauge invariant model. Consistently with result (A),
$\rho\approx 0$ over the line $q=0.5$ and only paramagnetic and
ferromagnetic phases are observed  with a transition around $p_c \approx
0.853$ for our example (see figure \ref{fig:nish_error}).  Result (B) is
only relevant if we can choose parameters such that $\xi_{ij}\ge 0$ (or
$p=1$) for any $(i,j)\in \Omega$, any random field $r_{ij}$ and any $B$ in
the Hamiltonian \ref{eq:hamilt_rot}. At Nishimori's condition, however,
$p=1$ implies that $\beta_N=\infty$ and $B_N=0$. Thus the model is a trivial
ferromagnet and result (B) is irrelevant. Yet if the inference model assumes
$\hat{p}<1$ while  $\xi_{ij}$ are generated with $p=1$, rigorous result (B)
forbids either a spin glass or a mixed phase to show up. Accordingly, figure
\ref{fig:ppHat_diag} exemplifies a phase diagram for $q>0.5$ with no mixed
phase for $p=1$ (true value) and $\hat{p}<1$ (estimated value).  Finally
result (C) implies in the absence of a mixed phase anywhere at the
Nishimori condition, which is indeed found as $\rho\approx 0$
\footnote{Actually we numerically find $\rho > 0$ over the same region of
distinctly long convergence times depicted in figure
\ref{fig:simul_convrate}. This, however, can be made arbitrarily small by
increasing the numerical precision employed.}.

This theoretical analysis may be repeated for the mismatched parameters case
introduced in the previous section, by setting $\beta = \alpha_{\hat{p}}$
and $B = \frac{\alpha_{\hat{q}}}{\alpha_{\hat{p}}}$ while considering $r_i$
and $\xi_i$ as quenched $\pm 1$ random variables with parameters $q$ and
$p$. Here again, the theoretical error obtained reproduces the empirical
one. In the $p-\hat{p}$ plane, however, a mixed phase appears for a region
of parameters. In this region, the free energy landscape becomes rugged, and
BP will hardly converge to its global minimum. Also the RS cavity analysis
does not necessarily provide asymptotically correct results, so the
thermodynamic quantities computed in this region may not reflect the actual
behavior of the system at equilibrium.

The onset of a mixed phase explains in part the degradation in algorithm's
performance for the upper left corner of the $p-\hat{p}$ plane. However,
empirical results show that convergence rates also worsens  outside this region
of parameters, as well as on the $p-q$ plane even at Nishimori's condition,
where no glassy phase is to be found. In the next section, we investigate
this issue further.
 
\section{Dynamical properties}
\label{sec:dynprop}

In order to understand such deterioration in the algorithm's performance, we
have studied the BP  dynamical system. From \ref{eq:BPeq}, we obtain

\begin{equation}
  \frac{\partial h_{i \to j}^{(t+1)}}{\partial h_{k \to i}^{(t)}} =
  \frac{\tanh(\beta) J_{ki}}{\cosh^2 (\beta h_{k \to i}^{(t)}) - \tanh^2
  (\beta) \sinh^2 (\beta h_{k\ \to i}^{(t)})} \mathbb{I} (k \in
  \partial i / j).
\end{equation}

The dynamical system in question has $n c$ equations, one for each direction
of each edge on the graph. In order to study the linear stability  of the BP
dynamical system, we  have calculated the spectral radius of the Jacobian
matrix evaluated at a fixed point, that is, $R = \max |\lambda
(\mathbb{J})|$, where $\mathbb{J}$ is a $n c \times n c$ matrix with entries
$\mathbb{J}_{ij,ki} = \frac{\partial h_{i \to j}^\ast}{\partial h_{k \to
i}^\ast}$.

\begin{figure}[ht]
 \centering
 \subfloat[Nishimori condition]{
   \includegraphics[width=0.37\textwidth]{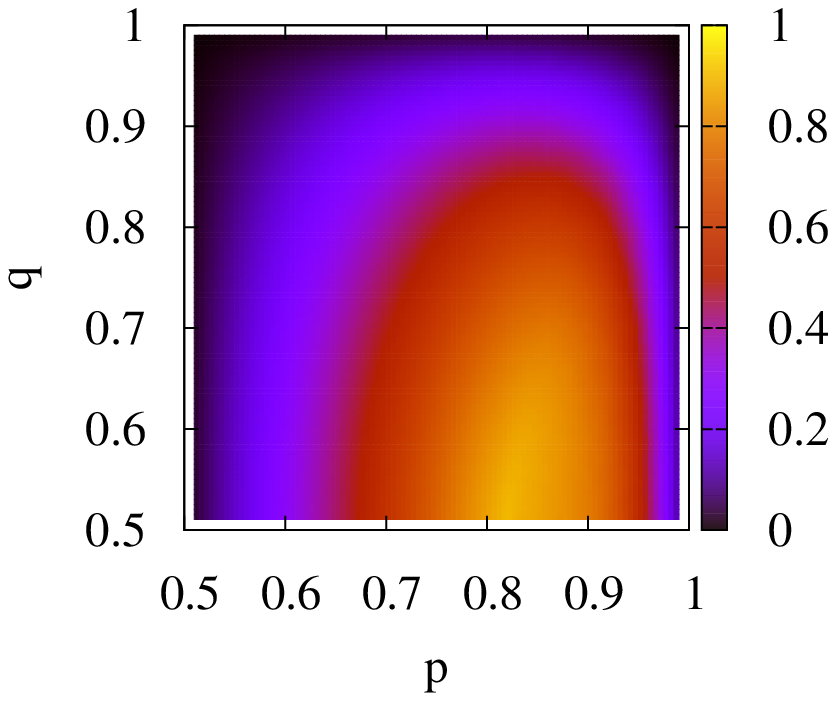}
 \label{fig:specrad_nish}
 }
 \subfloat[mismatched $p$]{
   \includegraphics[width=0.37\textwidth]{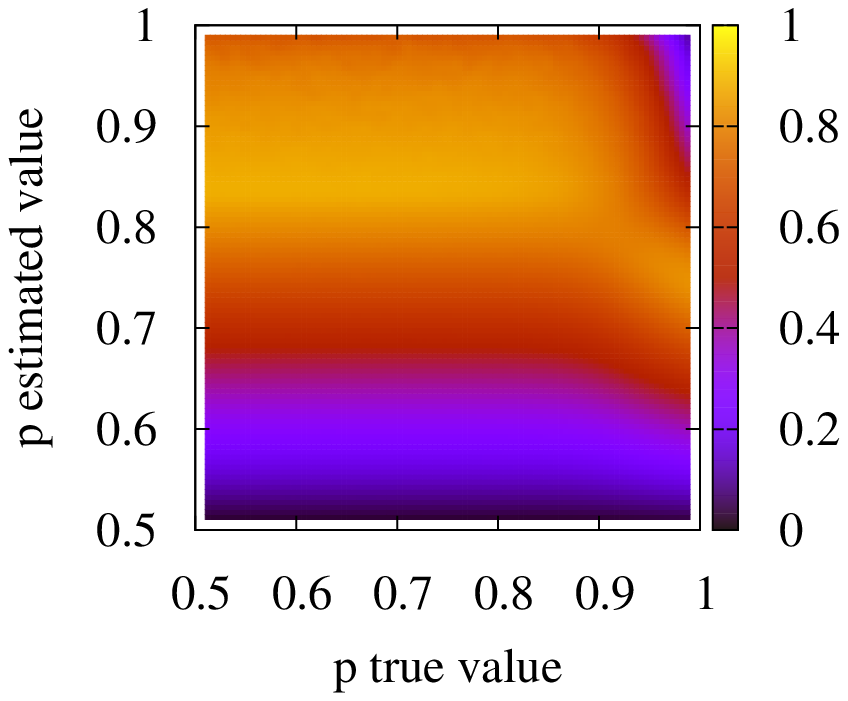}
 \label{fig:specrad_ppHat}
 }
 \caption{{\bf Average spectral radius} of the Jacobian of the BP dynamical
 system. For each value of $p$ and $q$ ($p$ and $\hat{p}$), the Jacobian's
 spectral radius was computed for 2500 runs of the algorithm and then
 averaged. Larger values of $R$ correlate with slower convergence times. 
Panel (a): at Nishimori's condition, to be compared  with \ref{fig:simul_convrate}. 
Panel (b): with a mismatch between $p$ and $\hat{p}$, to be compared  with
 \ref{fig:ppHat_convrate}. }
\label{fig:specrad}
\end{figure}

The spectral radius gives us a measure of the \emph{convergence rate}
of the algorithm. In fact, as figure \ref{fig:specrad} shows, regions where
the value of $R$ is larger coincide with those where the algorithm converges
more slowly, in average. 

Interestingly, this quantity may be also studied within the RS cavity
scheme. The population dynamics algorithm, which we have used to obtain
samples of $\prob(h)$, may be also seen as a dynamical system (the so
called, {\it density evolution} equations): 
\begin{equation}
  \left\{
  \begin{array}{l l}
   u_i^{(\ell)} &= \frac{1}{\beta} \tanh^{-1} \left[ \tanh (\beta \xi)
   \tanh (\beta h_i^{(\ell)}) \right],\\
   h_i^{(\ell+1)} &= Br + \sum_{j = 1}^{c-1} u_{\gamma(j)}^{(\ell)},\\
  \end{array} \right.
\end{equation}
with $r$ and $\xi$ independently sampled for each $i$, and $\gamma(\cdot)$
representing random indices, that is $\gamma(\cdot) \sim \textrm{Uniform} ([N])$. 

\begin{figure}[ht]
 \centering
 \includegraphics[width=0.6\textwidth]{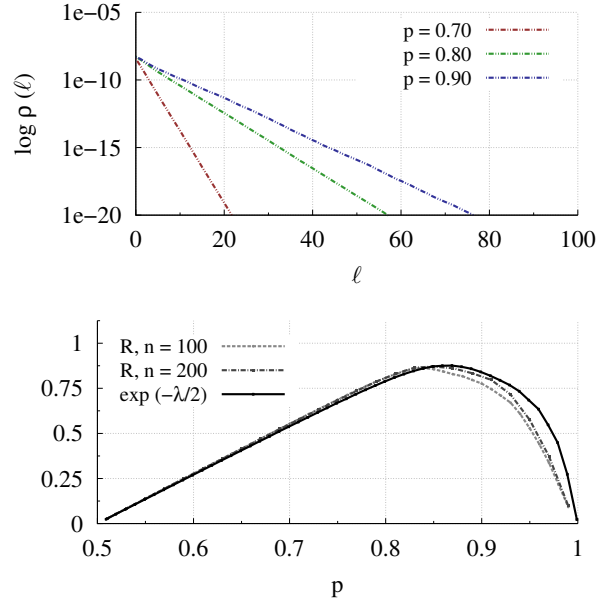}
 \caption{{\bf Decay of $\rho(\ell)$.} Top panel: using population dynamics  we find 
the rate  $\lambda$  of decay  as a function of $p$ for  $q = 0.6$. Bottom panel: the average spectral radius $R$ for 
finite size ($n=100,200$) maps are closely related  to  $e^{-\lambda/2}$. Notice the finite size effect in the large $p$ region. }
\label{fig:rho-t}
\end{figure}

In this context, the evolution of the order parameter $\rho$, $\rho (\ell) =
\frac{1}{N} \sum_{i = 1}^{N} ({u_i^\prime}^{(\ell)} - {u_i}^{(\ell)})^2$,
can be studied.  As it can be seen in figure \ref{fig:rho-t}, $\rho (\ell)$
decays exponentially at a constant rate $\lambda$, that is, $\rho \propto
e^{-\lambda \ell}$.  In the neighborhood of a fixed point the decay rate of
$\sqrt{\rho}$ is given by the dominant eigenvalue of the Jacobian matrix. We
thus expect the relation $R \propto  e^{-\frac{\lambda}{2}}$ to hold, where
$R$ is the spectral radius of the Jacobian matrix.  This relationship  is
clearly discerned in figure \ref{fig:rho-t} --- thus by computing the decay
rate $\lambda$ for $\rho$, we also learn about the algorithm convergence
rate.

We have observed two mechanisms leading to performance degradation: the
onset of a glassy phase and the decreased stability of the BP fixed point.
The former is a limitation intrinsic to the inference problem, the latter an
issue that probably could be addressed by modifying the approximate
inference algorithm. We however observe that the stability of the fixed
point decreases as the parameters approach a mixed phase, as $p = 0.8536, q
= 0.5$ defines a multicritical point in a model with $c=3$
\cite{Matsuda2011}, thus suggesting this can also be an intrinsic limitation
of the problem.

\section{Robustness}
\label{sec:robust}

To this point our analysis has only considered ratings $\{J_{ij}\}$
distributed over a regular random graph of fixed degree $c = 3$ and issued
exactly as assumed by the inference model. In this section, we relax these
assumptions to access both the performance of the algorithm and the validity
of the theoretical analysis under more general conditions.

\begin{figure}[ht]
 \centering
 \subfloat[empirical error]{
   \includegraphics[width=0.46\textwidth]{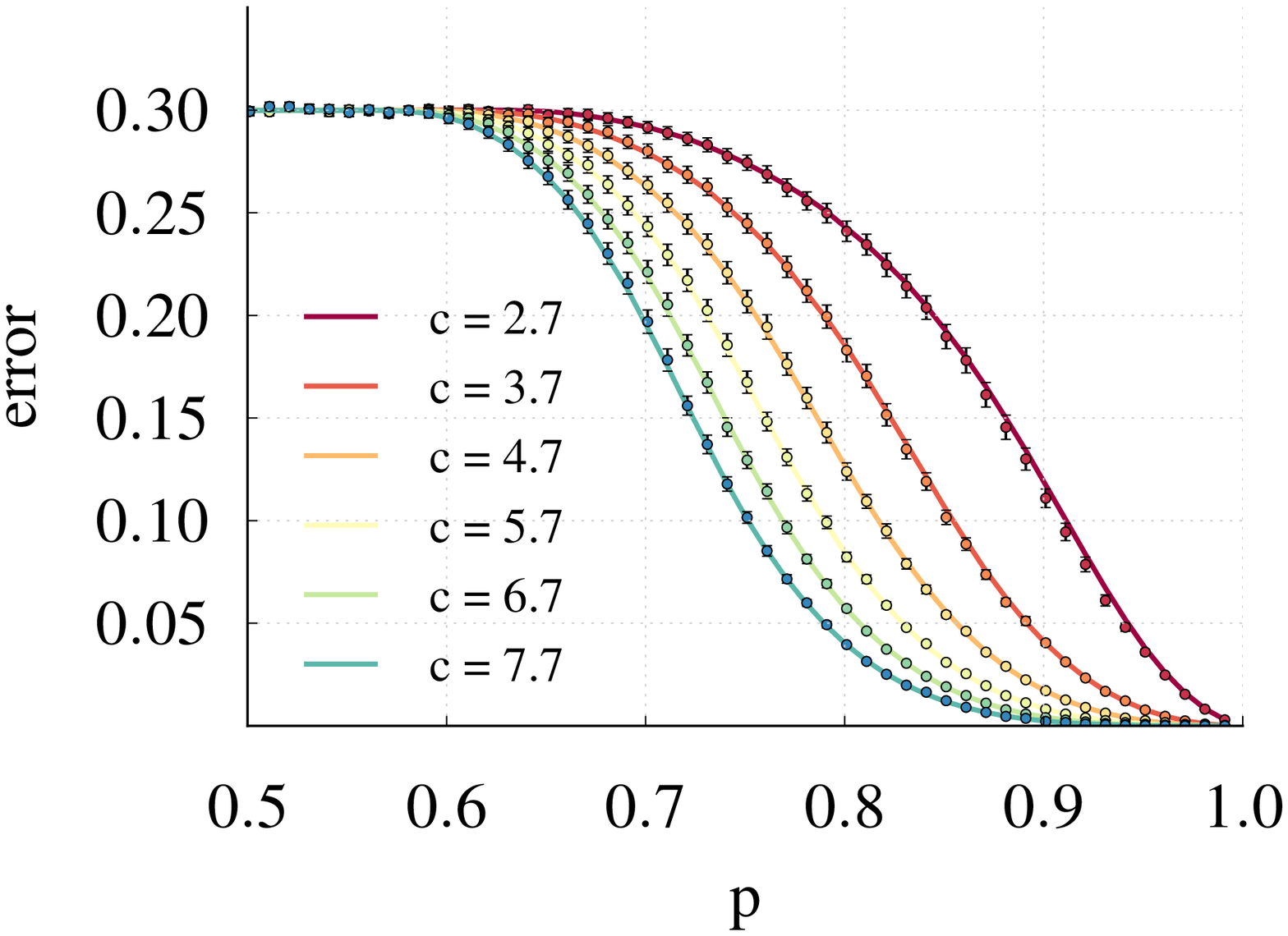}
 \label{fig:numNbr_error}
 }
 \subfloat[iterations to convergence]{
   \includegraphics[width=0.37\textwidth]{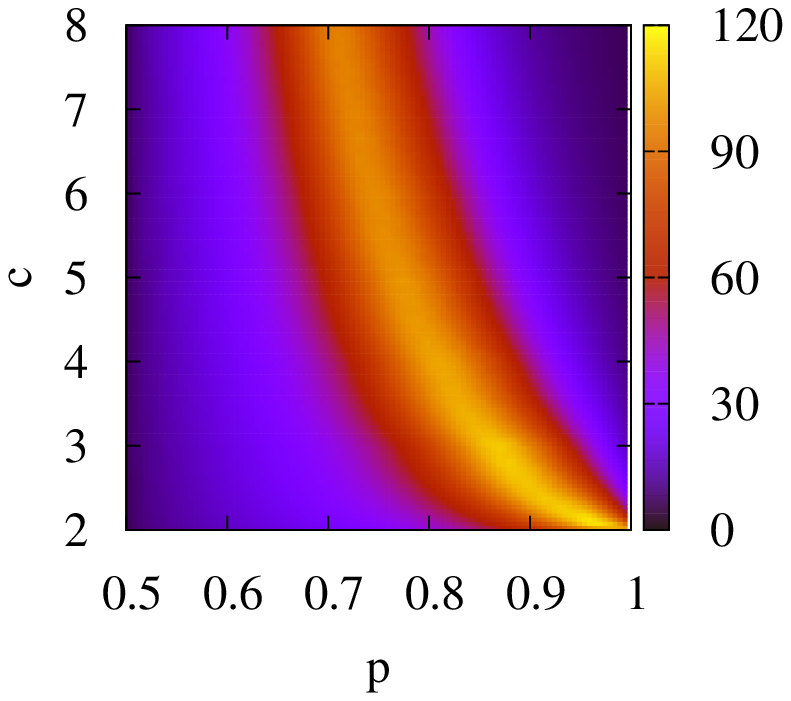}
 \label{fig:numNbr_convrate}
 }

 \vspace{0.2cm}

 \caption{{\bf Empirical and theoretical analysis at Nishimori's condition  for a random graph} with
 degree distribution given by eq.  \ref{eq:degprof} and $q = 0.7$. The
 parameter $c$ is the average node degree while $p$ is the signal level.
 Panel (a) compares  empirical (bars) and theoretical (lines) errors as a function of $p$ for (from top to bottom) $c=2.7$ to $c=7.7$ in steps of size $1$ 
 . Panel (b) depicts
 the average number of iterations for the BP algorithm to converge.}
\label{fig:numNbr}
\end{figure}

\subsection{Graph topologies}

The set $\Omega$ of ratings issued by network entities define a graph
$\mathcal{G}$ with each vertex representing an entity and an edge $(i,j)$
being connected if and only if $(i, j) \in \Omega$. The theoretical analysis
based on the replica-symmetric cavity equation \ref{eq:cav-rot_int} relies
on specifying an ensemble of graphs represented by a particular degree
distribution (or profile).  In the previous discussion we have used an
ensemble of regular random graphs with a degree profile given by $\Lambda_c
(\gamma) = \delta\,(\gamma - c)$. A natural extension to that is allowing
non-integer values of $c$, for that we introduce:
\begin{equation}
 \Lambda_c (\gamma) = \left\{
 \begin{array}{l l}
 1 - (c - \lfloor c\rfloor) & \quad \textrm{for $\gamma =
 \lfloor c\rfloor$,} \\
 c - \lfloor c\rfloor & \quad \textrm{for $\gamma =
 \lfloor c\rfloor + 1$,} \\
 0 & \quad \textrm{otherwise.}
 \end{array} \right.
 \label{eq:degprof}
\end{equation}
In this way for $c = 2.3$ we would have 30\% of the nodes with degree 3 and
the remaining with degree 2.

\begin{figure}[ht]
 \centering
 \includegraphics[width=0.57\textwidth]{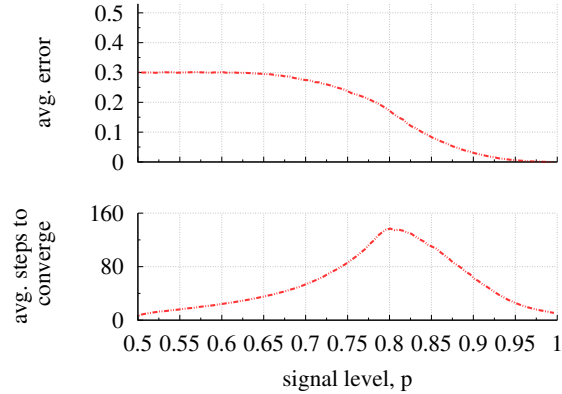}
 \caption{{\bf Simulations in a square lattice}, for $q = 0.7$ and different values of $p$. Despite the
 short cycle lengths, the algorithm exhibits a good performance.}
 \label{fig:square}
\end{figure}

The empirical analysis is done by simulating instances sampled from this
ensemble of graphs.  Figure \ref{fig:numNbr}(a) compares empirical and
theoretical errors at Nishimori's condition as a function of $p$ for
several $c$ values, $q=0.7$.   Figure \ref{fig:numNbr}(b) depicts the
average number of iterations to convergence as a function of $p$ and $c$
which is explained by the stability of the unique BP fixed point. 

The BP algorithm calculates exact marginals and allows for optimal Bayesian
inference when the subjacent graph is a tree.  The performance of the BP
algorithm, however, can be studied by simulation on any topology. Figure
\ref{fig:square} shows the resulting performance measures as a function of
$p$ for $q=0.7$ and for $\mathcal{G}$  chosen to be a square lattice in two
dimensions. The average error is always smaller than $1-q$ and vanishes as
$p\rightarrow 1$  showing that even in this case the BP algorithm may yield
good results. 

\subsection{Attacks}
\label{sec:attacks}

Malicious entities may issue ill-intentioned ratings to trick the reputation
system and malfunctioning devices may issue erroneous ratings. These
scenarios of targeted attacks can also be considered by the inference model
and studied within the same theoretical framework and simulation techniques.

We suppose that a fraction $f$ of the ratings are issued by noisy
entities while the inference process remains unchanged. Lets call the subset
of noisy ratings $\Omega_f\subset\Omega$. To simulate this scenario, we
uniformly sample and fix a fraction $\frac{f}{2}$ of the entities to be
noisy, issuing ratings as $J_{ij} = \eta_{ij}$, where $\eta_{ij} =
\eta_{ji}$ is a $\pm 1$ random variable with parameter $z = \frac{1}{2}$
and $(i,j) \in \Omega_f$ --- since we require $J_{ji} = J_{ij}$, the total
fractions of noisy ratings will be $f$. We then run the BP algorithm and
study how the performance changes with $f$.

\begin{figure}[ht]
 \centering
 \includegraphics[width=0.37\textwidth]{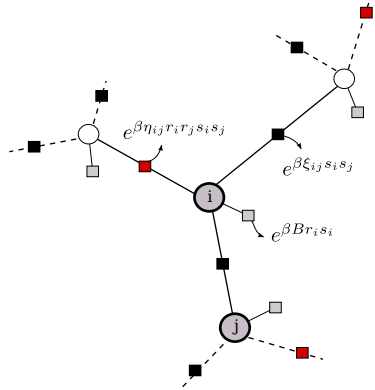}
 \caption{{\bf Factor graph for the posterior \ref{eq:attackpost}.}
 A fraction $f$ of the function nodes represent noisy ratings (red
 squares).}
\label{fig:factor_att}
\end{figure}

For the theoretical analysis we calculate performance measures averaged
over every disorder component: regular random graph with $c=3$ in our
example, symmetric communication noise $\bm{\xi}$, symmetric random ratings
$\bm{\eta}$ and symmetric ratings $\bm{J}$. As we are interested in checking
algorithm robustness, we also assume that the inference scheme  has no
knowledge that  the reputation system is under attack.

We first rewrite the posterior by taking into account noisy ratings: 
\begin{equation} 
  \prob(\bm{r} | J) \propto \prod_{(i, j) \in \Omega \setminus \Omega_f}
  \prob(\xi_{ij}) \prod_{(i, j) \in \Omega_f} \prob(\eta_{ij}) \prod_i
  \prob(r_i).
\end{equation} 

Following the previous steps yields:
  \begin{equation}
    \prob(\bm{r} | J) \propto e^{\alpha_p  \sum_{(i, j) \in \Omega \setminus\Omega_f} J_{ij} r_i r_j }
                            e^  {\alpha_z  \sum_{(i, j) \in \Omega_f} \eta_{ij} + \alpha_q \sum_i r_i }
  \label{eq:attackpost}
  \end{equation}

Since the algorithm considers the ratings $\Omega_f$ as subject to the same
communication noise as regular ratings, we have $\alpha_z = \alpha_p$.
The gauge transformed Hamiltonian for an equilibrium statistical mechanics
description is
\begin{equation}
 H (\bm{s}) = -\sum_{(i, j) \in \Omega \setminus \Omega_f} \xi_{ij}
 s_i s_j -\sum_{(i, j) \in \Omega_f} \eta_{ij} r_i r_j s_i s_j
 -B \sum_i r_i s_i.
 \label{eq:hamilt-att}
\end{equation}

\begin{figure}[ht]
 \centering
 \subfloat[empirical error]{
   \includegraphics[width=0.37\textwidth]{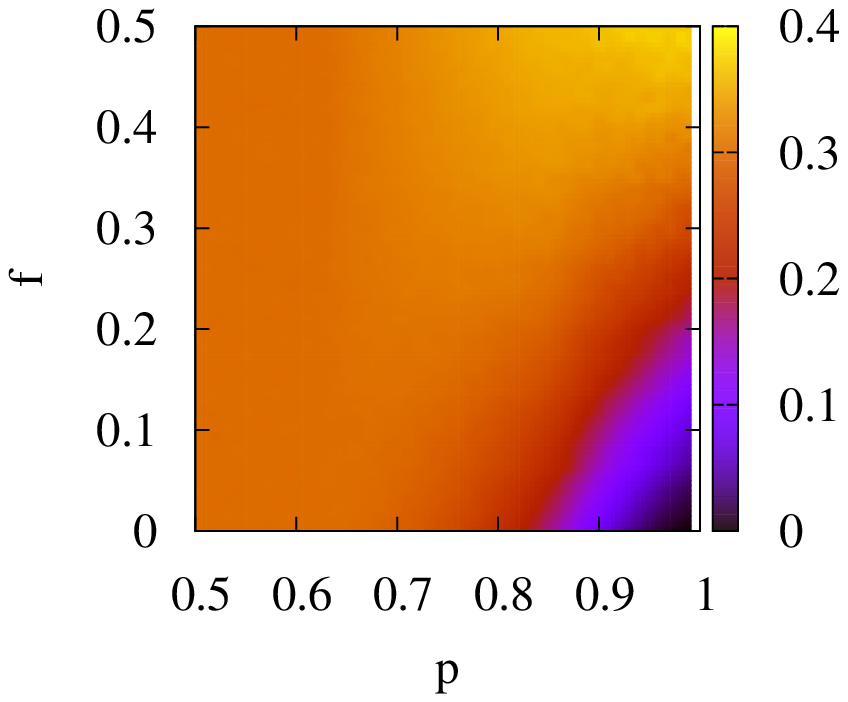}
 \label{fig:attacks_error}
 }
 \subfloat[iterations to convergence]{
   \includegraphics[width=0.37\textwidth]{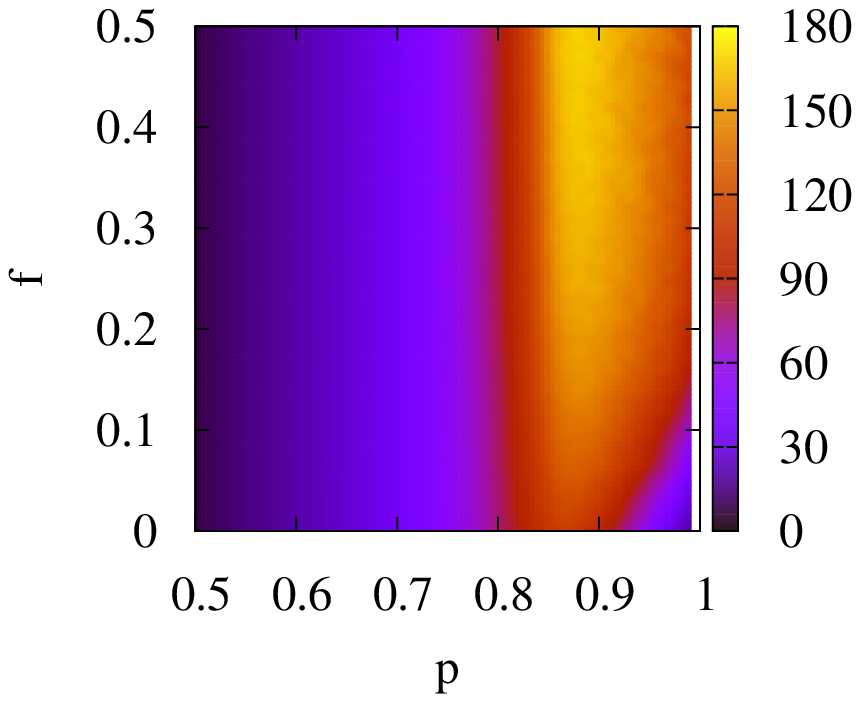}
 \label{fig:attacks_convrate}
 }
 \vspace{0.2cm}

 \subfloat[theoretical error]{
   \includegraphics[width=0.37\textwidth]{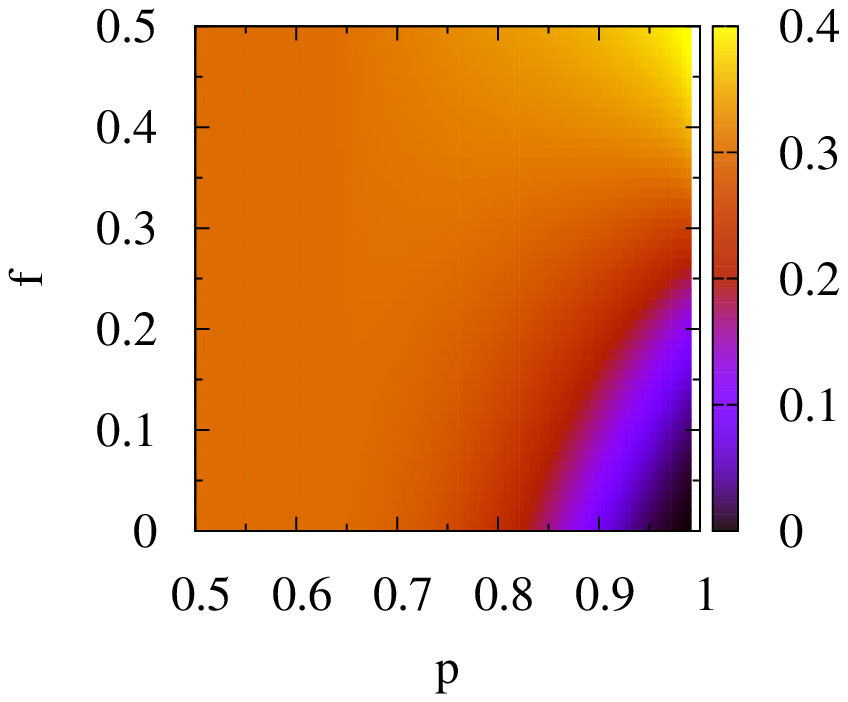}
 \label{fig:attacks_therror}
 }
 \subfloat[phase diagram]{
   \includegraphics[width=0.37\textwidth]{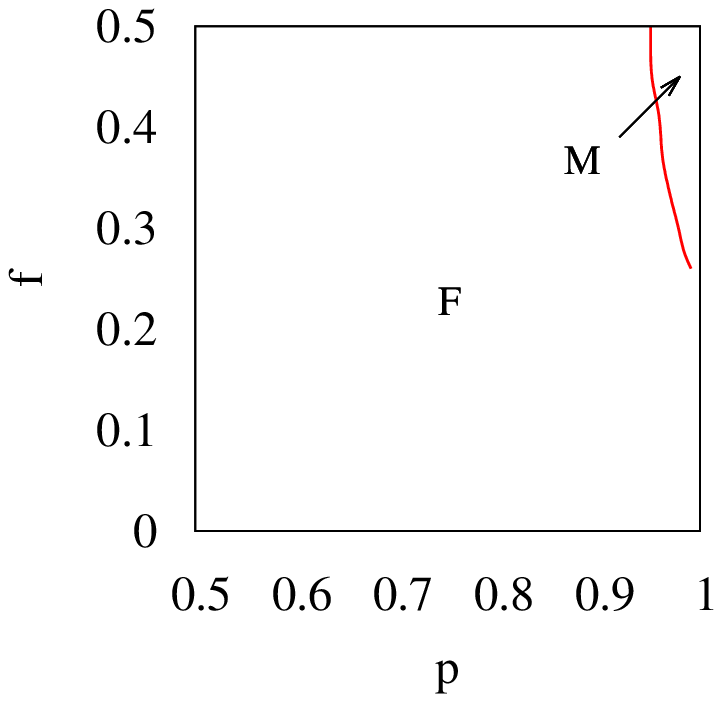}
 \label{fig:attacks_diag}
 }
 \caption{{\bf Fraction $f$ of nodes broadcasting random ratings.} Panels (a) and
 (b) depict simulation results for the error and for the average number of
 iterations to convergence. Panels (c) and (d) show theoretical error and
 phase diagram using the replica symmetric cavity approach. As expected, the
 error increases with $f$.}
 \label{fig:attacks}
\end{figure}

There is a new term $-\sum_{(i, j) \in \Omega_f} \eta_{ij} r_i r_j s_i s_j$
in the Hamiltonian, which can be treated by the inclusion of a new type of
function node to the factor graph representation for the posterior
\ref{eq:attackpost}. Figure \ref{fig:factor_att} provides a snapshot of
this factor graph. The replica symmetric cavity description can then be
written as
\begin{equation}
 h \stackrel{d}{=} \lld Br + \sum_{i=1}^{c-1} u_i (h) \rrd_{r, \xi, \eta},
 \label{eq:cav-att}
\end{equation}
where $u_i (h) = \frac{1}{\beta} \tanh^{-1} \left[ \tanh (\beta \xi_{i})
\tanh (\beta h_i) \right]$ with probability $1-f$, and with probability
$f$
\begin{equation*}
  u_i(h) = \frac{1}{\beta} \tanh^{-1} \left[ \tanh (\beta \eta_{i} r r_{i})
    \tanh (\beta h_i) \right],
\end{equation*}
where $\eta_i$ is a $\pm 1$ random variable with $\prob(\eta_i = 1) =
\frac{1}{2}$, $r$ is the same used in eq. \ref{eq:cav-att} and the $\{r_i\}$
are independently sampled from $r$.

Figure \ref{fig:attacks} depicts simulation results for the error in Panel
(a) and for the number of iterations to convergence in Panel (b). Panels
(c) and (d) show theoretical error and phase diagram using the replica
symmetric cavity approach. The increased time to convergence inside the
ferromagnetic phase is explained  by the decreased  stability of the still
unique BP fixed point. Inside the mixed phase new fixed points emerge and
the inference process is fundamentally faulty as can be seen in the top
right corner of the error surface.   

\section{Conclusions}
\label{sec:concl}

The use of a belief propagation (BP) algorithm for approximate inference in
reputation systems has been introduced in \cite{Ermon2009b,Ayday2011}. We
here extend previous work by calculating performance measures using the
replica symmetric cavity approach after expressing the inference problem in
terms of equilibrium statistical mechanics. We apply this framework to a
basic scenario and to three simple variations.

The framework is very general and allows for the study of the algorithm
performance when subjected to several scenarios of practical interest such
as the presence of collusions, parameter mismatches and  targeted attacks.
Other questions of practical interest remain. Algorithms based on BP
approximate inference seem to represent an interesting alternative for
reputation systems such as wireless sensor networks, however implementation
details that have been purposefully ignored in our analysis certainly
deserve a more thorough analysis. For instance, sensors often operate with
very limited resources, so that sampling of ratings and running of the algorithm
should be scheduled taking these limitations into account. Also faulty elements would behave 
differently with lower signal to noise rates. In another direction, in a
distributed scheme it would be interesting to study the role of different prescriptions for 
the matrix $\bm{J}$. 

From a theoretical point of view reinforced belief propagation or survey
propagation techniques promise better results in the deteriorated
performance glassy phase. Also, expectation maximization-belief propagation
\cite{Decelle2011} could allow the algorithm to run without the need of
supplying signal level $p$ and reputation bias $q$ as inputs. For scenarios
involving targeted attacks more information could be built into the rating
mechanism that may allow for improved inference algorithms.  

\ack{
This work was supported by CNPq, FAPESP under grant 2012/12363-8, and The
Center for Natural and Artificial Information Processing Systems of the
University of São Paulo (CNAIPS-USP).
}

\appendix
\section{Computation of marginals using belief propagation}
 \label{sec:mpm} 

The algorithm takes as input a distribution $\prob(h)$ from which the
messages are initially sampled, a maximum to the number of iterations
$t_{\max}$, a precision $\epsilon$ for convergence and estimated values of
$p$ and $q$, $\{\hat{p}, \hat{q}\}$. In what follows, we have used
$P^{(0)} (h) = \delta (h)$, $t_{\max} \sim 100$, $\epsilon \sim 10^{-7}$
and $\{\hat{p}, \hat{q}\} = \{p, q\}$ --- this last condition is later
relaxed. The complete pseudocode is as follows:

 \vspace*{0.5cm}
 \algsetup{indent=1em}
 \begin{algorithmic}[1]
 \REQUIRE {$P^{(0)} (h)$, $t_{\max}$, $\epsilon$, $\{\hat{p},
 \hat{q}\}$; $\{J_{ij}\}$, $\mathcal{G}$}
 \ENSURE {$\{\hat{r}_i\}$}
 \STATE initialize $\{h_{i \to j}\}$ sampling from $P^{(0)} (h)$
 \STATE $\beta \gets \alpha_{\hat{p}}$, $B \gets {\alpha_{\hat{q}}} / {\alpha_{\hat{p}}}$

 \WHILE{$\Delta \geq \epsilon$ \AND $t < t_{\max}$}
 \FOR{$i = 1 \to n$, $j \in \nb{i}$}
 \STATE $h_{i \to j}' = B + \sum_{k \in \nb{i} / j} u_{k \to i} (J_{ki}, h_{k \to i})$
 \ENDFOR
 \STATE $\Delta \gets \max\, |h_{i \to j}' - h_{i \to j}|$
 \STATE $t \gets t+1$, $\{h_{i \to j}\} \gets \{h_{i \to j}'\}$
 \ENDWHILE
 \IF{$\Delta < \epsilon$}
 \FOR{$i = 1$ \TO $n$}
 \STATE $\hat{h}_i = B + \sum_{k \in \nb{i}} u_{k \to i} (J_{ki}, h_{k \to i})$
 \STATE $\hat{r}_i = \sgn (\hat{h}_i)$
 \ENDFOR
 \ENDIF
 \end{algorithmic}
 \vspace{0.3cm}

\section{Population dynamics}
\label{sec:popdyn}

The population dynamics algorithm provides an approximate solution to
\ref{eq:cav-rot} by iterating

\begin{equation}
 \left\{
 \begin{array}{l l}
 u_i^{(\ell)} &= \frac{1}{\beta} \tanh^{-1} \left[ \tanh (\beta \xi)
 \tanh (\beta h_i^{(\ell)}) \right],\\
 h_i^{(\ell+1)} &= Br + \sum_{j = 1}^{c-1} u_j^{(\ell)}.\\
 \end{array} \right.
 \label{eq:updt}
\end{equation}

We introduce two arrays of length $N = 10^4$: $\bm{h}$ and $\bm{u}$. At the
first step, the elements of $\bm{u}$ are initialized. We have considered two
possible ways of initializing $\bm{u}$: by uniformly sampling from
$[-\varepsilon, \varepsilon]$, $\varepsilon = 10^{-2}$, or by assigning
$u_i^{(0)} = \xi_i$ (i.e., $\tanh (\beta h_i^{(0)}) = 1$); the results
obtained in our analysis were very similar for both. For discussions
regarding the use of different initial conditions, the reader may refer to
\cite{Zdeborova2009,Matsuda2011}. 

Next, the elements of $\bm{h}$ are updated according to the rule, with
$\{u_j\}$ uniformly sampled from $\bm{u}$ and $r$ sampled from $\prob(r)$;
and the elements of $\bm{u}$ are calculated from the respective element in
$\bm{h}$ and $\xi$ sampled from $\prob(\xi)$. The process is repeated $\tau
= 5000$ times. After this large number of iterations, the array $\bm{h}$
should be approximately distributed as the real distribution $\prob(h)$,
and we are then able to calculate the desired averages.

In order to calculate $m = \left\langle \tanh (\beta \hat{h})
\right\rangle_{\hat{h}}$, we may introduce an array $\bm{\hat{h}}$ with the
same length $N$, and since $\hat{h}$ is given by a sum with an extra term,
the array elements are computed by simply summing the elements of $\bm{h}$
with some uniformly sampled element of $\bm{u}$. We then have $m \approx
\frac{1}{N} \sum_i \tanh (\bm{\hat{h}} [i])$.

\section*{References}
\bibliographystyle{unsrt}
\bibliography{refs}

\end{document}